\documentclass[11pt,preprint]{aastex}

\begin{document}


\title{LITHIUM AND BERYLLIUM IN ONE-SOLAR-MASS STARS}

\author{Ann Merchant Boesgaard\altaffilmark{1}}
\affil{Institute for Astronomy, University of Hawai`i at M\-anoa, \\ 2680
Woodlawn Drive, Honolulu, HI {\ \ }96822 \\ } 

\email{annmb@hawaii.edu}

\author{Constantine~P.~Deliyannis\altaffilmark{1}}
\affil{Department of Astronomy, Indiana University, 727 East 3rd Street, \\
Swain Hall West 319, Bloomington, IN {\ \ }47405-7105 \\ }

\email{cdeliyan@indiana.edu}

\author{Michael G.~Lum\altaffilmark{1}}
\affil{Institute for Astronomy, University of Hawai`i at M\-anoa, \\ 2680
Woodlawn Drive, Honolulu, HI {\ \ }96822 \\ } 

\email{mikelum@mac.com}

\author{Ashley Chontos\altaffilmark{1}} \affil{Institute for Astronomy,
University of Hawai`i at M\-anoa, \\ 2680 Woodlawn Drive, Honolulu, HI {\ \
}96822 \\ } \affil{Astrophysical Sciences, Peyton Hall, Princeton University,
Princeton, NJ {\ \ }08544 \\ }

\email{ashleychontos@princeton.edu}

\altaffiltext{1}{Visiting Astronomer, W.~M.~Keck Observatory jointly operated
 by the California Institute of Technology and the University of California.}

\begin{abstract}
The surface content of lithium (Li) and beryllium (Be) in stars can reveal
important information about the temperature structure and physical processes
in their interior regions.  This study focuses on solar-type stars with a
sample that is more precisely defined than done previously.  Our selection of
stars studied for Be is constrained by five parameters: mass, temperature,
surface gravity, metallicity, and age to be similar to the Sun and is focussed
on stars within $\pm$0.02 of 1 M$_{\odot}$.  We have used the Keck I telescope
with HIRES to obtain spectra of the Be II spectral region of 52 such stars at
high spectral resolution ($\sim$45,000) and high signal-to-noise ratios.
While the spread in Li in these stars is greater than a factor of 400, the
spread in Be is only 2.7 times.  Two stars were without any Be, perhaps due to
a merger or a mass transfer with a companion.  We find a steep trend of Li
with temperature but little for Be.  While there is a downward trend in Li
with [Fe/H] from $-$0.4 to +0.4 due to stellar depletion, there is a small
increase in Be with Fe from Galactic Be enrichment.  While there is a broad
decline in Li with age, there may be a small increase in Be with age, though
age is less well-determined.  In the subset of stars closest to the Sun in
temperature and other parameters we find that the ratio of the abundances of
Be to Li is much lower than predicted by models; there may be other mixing
mechanisms causing additional Li depletion.
\end{abstract}

\section{INTRODUCTION}

In 1951 Greenstein \& Richardson (1951) showed that Li in the Sun was only
one-hundredth as abundant as on Earth.  They speculated that if material
containing Li were circulated to temperatures near 3 x 10$^6$ K, the Li nuclei
would be lost by thermonuclear disintegration.  They suggested that any mixing
of the surface material with regions of such temperatures would have to be
very slow for there to be any surface Li at all.  The most recent assessment
of solar Li is A(Li) = 0.96 $\pm$0.06, where A(Li) = log N(Li) - log N(H) 
+12.00 (Asplund, Amarsi \& Grevesse (2021).  The solar system Li content is
A(Li) = 3.27 $\pm$0.03 (Lodders (2021) or about 200 times greater.

By 1954 Greenstein \& Tandberg Hansen (1954) measured the abundance of Be in
the Sun from lines of both Be II and Be I and found it to be similar to the Be
content in meteorites and on Earth.  They knew then that the Sun had not
destroyed its Be and thus Be nuclei would not have been transported to
temperatures of 3.6 x 10$^6$ K.  The current evaluation of solar Be is A(Be) =
1.38 $\pm$0.09 (Asplund, Amarsi \& Grevesse 2021) and of meteoritic Be is 1.31
$\pm$0.04 (Lodders 2021), where A(Be) = log N(Be) - log N(H) +12.00.

 The solar Li abundance is observed to be down by more than 200 compared to
the solar system value while the solar and meteoritic Be abundances are
comparable to each other.  Standard Stellar Evolution Theory (SSET) has been
unable to account for this low value of solar Li, while recognizing that any
main-sequence mixing processes had to be slow and involve additional mixing
mechanisms.  Somers \& Pinsonneault (2016) discuss Li depletion with a
modification to the role of differential rotation.  They include more angular
momentum transport and determine that core-envelope recoupling produces
efficient mixing.

In order to understand the solar Li depletion, several studies have been done
on Li in solar twins, e.g. Matsuno, Aoki, Suda et al.~ (2017), Th\'evenin,
Oreshina, Baturin et al.~(2017).  A recent example is that of Carlos,
Melendez, Spina et al.~(2019) who determined Li abundances in 77 solar
``twins.''  They conclude that the Sun ``is unusually Li-deficient for it's
age.''  They define twins by temperature ($\pm$100 K), log g ($\pm$0.01) and
[Fe/H] ($\pm$0.01), approximately.  Their sample does not constrain mass or
age, but 30 stars are within $\pm$2\% of the solar mass and of those 16 are
within 2 billion years of the age of the Sun.  For those 16 their values for
LTE Li range from 0.91 to 1.76 plus one with an upper limit of $<$0.49.  The
Sun at 0.96 is close to the bottom of the range.

Boesgaard, Lum \& Deliyannis (2020) studied Li and Be in the
\underline{solar-age} and \underline{solar-metallicity} open cluster, M 67.
In their Figures 7 and 8 they indicated the position of the solar Li value.
In both figures of Li vs.~temperature and Li vs.~mass the Sun is in the lower
third of the range of Li in M 67, but not particularly out of line in the Li
trends.

Although it is much more difficult to make observations of Be than Li, its
abundance is an important ingredient to our understanding of stellar
interiors.  Both Li and Be are destroyed by primarily by energetic protons but
whereas Li survives to a depth where T is $\sim$2.5 x 10$^{6}$ K, Be nuclei
survive deeper in the star to $\sim$3.5 x 10$^{6}$ K.  Abundances of both
elements, therefore, produce information about the interiors and the
processes at work that affect their depletion.

There have been some previous attempts to find Be in solar-like stars with
known Li.  Boesgaard \& Hollek (2009) found Be in 50 ``solar mass'' stars,
although their sample ranged in mass from 0.90 to 1.09 M$_{\odot}$.  Most of
their stars were below [Fe/H] = $-$0.15 and the Be abundances in the thin disk
stars fit well with the Be-Fe trend found for halo metal-poor stars found by
Boesgaard, Rich, Levesque et al.~(2011); that relationship was shown to extend
over 4 orders of magnitude in [Fe/H] and 3 orders of magnitude in A(Be) as
shown in their Figure 11.  Four stars in their sample, all subgiants showed
strong depletions of Be.

\begin{figure}[h]
\epsscale{0.7}
\plotone{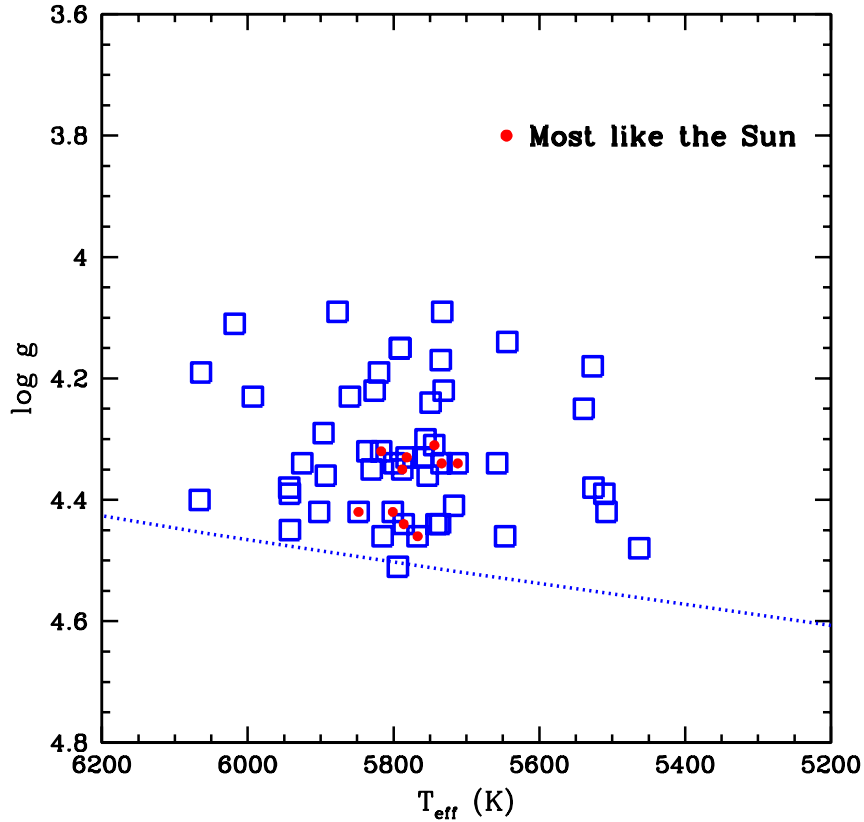}
\caption{The distribution of the stars we observed by their surface
temperatures and gravities as log g from Ramirez et al.~(2012), shown as blue
open squares.  The zero-age main sequence is indicated by the dotted line from
Demarque, Woo, Kim et al.~(2004).  The ten stars that are closest to the Sun
in all five parameters have small red dots inside the open squares.}
\end{figure}

Takeda, Tajitsu, Honda et al.~(2011) determined Be abundance in 118 solar
``analogs'' but had an even larger range in parameters with [Fe/H] from $-$0.6
to +0.4 and temperatures from 5000 to 7000 K, but centered on 5775 K, and
masses of 0.8 to 1.7 M$_{\odot}$.  They found little dependence of A(Be) on
metallicity, temperature or age in the early G stars, but a small increase
with v sin i between 1 and 7 km sec$^{-1}$.  In their sample there were four
stars that were severely deficient in Be and Li.

Chmielewski et al.~(1975) did a careful examination of the solar Be amount and
found A(Be) = 1.15 $\pm$0.20.  This value is lower by almost a factor of two
than the meteoritic value of 1.42 (Anders \& Grevesse (1989).  Balachandran \&
Bell (1998) suggest this deficiency is not real, but is due to missing uv
opacity or differences between the predicted and observed solar uv flux.
However, as mentioned above, recent compilations of solar and solar system
abundances show better agreement for Be.  Asplund, Amarsi \& Grevesse (2021)
found A(Be) = 1.38 $\pm$0.09 for the Sun.  For meteorites Lodders (2022) gives
A(Be) = 1.31 $\pm$0.04.  

In this work we take another observational approach to understand Li and Be in
stars and to help clarify the solar Li and Be issues.  The sample presented in
this paper is more restrictive than those done previously.  It has a narrower
range in mass, temperature and [Fe/H].  All of our targets are from a
collection of 1581 FGK stars with a consistent set of stellar parameters and
Li abundances by Ramirez, Fish, Lambert et al.~(2012).  They have determined
and cataloged their Li abundances, ages, effective temperatures, surface
gravities, masses, and metallicities along with the associated errors in those
determinations.  They also indicate whether each star belongs to the thin
disc, thick disc or halo population.  From their list we have selected stars
to observe for Be that are predominently within $\pm$2\% of 1 M$_{\odot}$.

\section{OBSERVATIONS AND STELLAR PARAMETERS}

We have observed Be in stars of one solar mass selected from the sample of
1381 FGK stars for which Ramirez et al.~(2012) determined Li abundances.
Their sample includes Li abundances in 671 newly observed stars supplemented
by 710 Li determinations by others normalized to their scale.  They give the
offsets they used for each of the seven other studies they included in their
Table 2.  For all of the stars they give errors in the stellar parameters.
For the solar mass stars we have observed the errors in the mass
determinations are given as +0.02, $-$0.03 or +0.03, $-$0.02.  (Two of our
stars, HIP 394 and HIP 42723, are out of our mass range and error range at
1.10 +0.15, $-$0.05 and 1.17 +0.17, respectively.)  All of our sample of stars
are from the thin disk, except HIP 35599 from the thick disk.  One star is a
subgiant, HIP 394.

\begin{figure}[h]
\epsscale{0.7}
\plotone{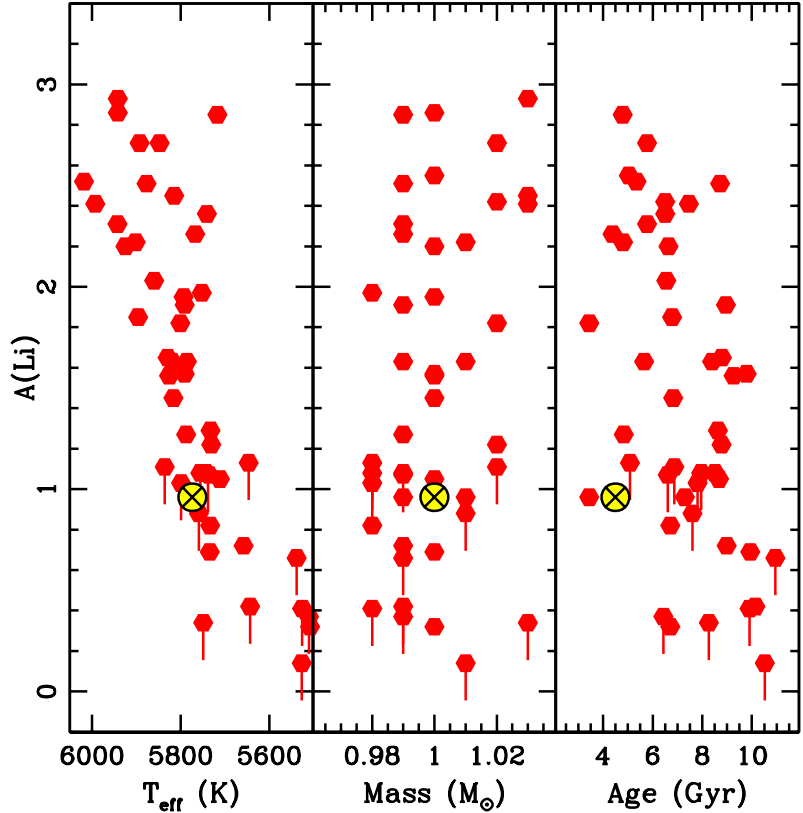}
\caption{Abundances of Li as A(Li) in the solar-mass stars we have observed
for Be shown as functions of surface temperature, stellar mass and age.  The
values are from Ramirez eta l.~(2012).  The position of the Sun is the yellow
circled X at A(Li) = 0.96 (Asplund, Amarsi \& Grevesse 2021).  The vertical
lines attached to some of the Li points indicate that those Li values are
upper limits.  The mass range is very close to solar, within $\pm$0.02.  The
solar temperature is 5772 K; most of our stars are within $\pm$100 K of that.
The age of the Sun is 4.60 Gyr; 75\% of our stars are within a factor of 2 of
that age.}
\end{figure}

Our Be spectra were obtained with HIRES (Vogt et al.~1994) on the Keck I
telescope with the 2004 upgraded version during the course of four observing
runs between January, 2014 and December, 2019.  The spectrograph has three ccd
chips and our wavelength coverage was from 3035 - 5880 \AA.  The Be II
resonance lines are at 3130.42 and 3131.06 \AA{}, on the blue chip near the
atmospheric cutoff $\sim$3000 \AA.  This blue ccd has a quantum efficiency of
94\% at 3130 \AA{} so we could observe our stars with short exposure times yet
high signal-to-noise ratios (S/N).  The spectral resolution was $\sim$45,000
or 0.023 \AA{} per pixel.

For our sample of 53 solar mass stars the exposures ranged from 1 - 20 minutes
with most being less than 10 minutes.  The values of S/N near the Be II lines
ranged from 50 to 190 with most between 70 - 100.  The stars were bright with
only six having V magnitudes fainter than 8.5.  In order to minimize the
effects of atmospheric attenuation and refraction we observed the stars as
close to the meridian as possible.  Our candidate list was long enough to make
this straightforward.

During each night of observing we obtained 1 s exposures of a Th-Ar comparison
spectrum at the beginning and end of the night for wavelength calibration.
Several exposures were taken of a quartz lamp to enable the flat-fielding of
the science frames.  These exposures for the blue ccd had to be 45 s to get
enough signal near 3130 \AA.  Typically 11 bias frames of 0$^s$ were taken for
background corrections.  The data reduction process was enabled by the MAKEE
pipeline (Barlow 2008).  The Th-Ar spectra turned out to be identical at the
beginning and end of the night and were used to make the preliminary
wavelength adjustments; the final wavelength corrections were done with IRAF
(Tody 1986, 1994)$\footnote{IRAF is distributed by the National Optical
Astronomy Observatories, which are operated by The Association of Universities
for Research in Astronomy, Inc. now NOIRLab, under cooperative agreement with
the National Science Foundation.}$.  We have on hand a Keck/HIRES spectrum
taken of the daytime sky at sunrise to use as a surrogate for the solar
spectrum in the Be II spectral region.  That exposure time was 20 min yielding
a S/N ratio of 138 (Boesgaard \& King 2002).

Table 1 gives a list of the stars observed for this program by HIP number and
by HD number.  The V magnitude, the UT date of the observation, the exposure
time in minutes and the resulting S/N of the reduced spectrum near the Be II
lines are listed.  We have used the stellar parameters from the uniform set
presented by Ramirez et al.~(2012).  Table 2 gives our sample of one solar
mass stars with their values for temperature ($T_{\rm eff}$), log g, [Fe/H],
microturbulent velocity, $\xi$, (calculated from the equation of Edvardsson et
al.~1993), the Li abundance, A(Li) = log N(Li)/log N(H) + 12.00, stellar mass
and age.

Figure 1 shows the distribution of the stars in Table 2 with $T_{\rm eff}$ and
log g along with a zero-age main sequence.  This shows that our stars are
somewhat evolved off the main sequence making their ages more reliable as
discussed by Ramirez et al.~(2012).  (One star, HIP 394, a subgiant with log g
= 3.76 is not plotted.)  The closest matches to the Sun are indicated with red
dots; see discussion in section 4.

Figure 2 plots the values for A(Li) in our Be sample as functions of
temperature, mass, and age.  (Again HIP 394 is not included.)  The position of
the Sun is given with the value of 0.96 $\pm$0.06 of Asplund, Amarsi \&
Grevesse (2021).  The range in stellar parameters of our observed stars is
vary narrow, but the Li abundances spread over three orders of magnitude.  For
the sample of stars we have observed, the position of the solar Li is in the
lower third of all of those parameters for stars in its cohort.

\section{ANALYSIS}
\smallskip
We have used the set of consistently determined stellar parameters of Ramirez
et al.~(2012) that are given in Table 2 to find Be abundances with our
Keck/HIRES spectra.  In order to determine the Be abundances we have used the
spectrum analysis program
MOOG$\footnote{http://www.as.utexas.edu/~chris/moog.html}$ (Sneden 1973,
Sneden et al.~2012 ) as updated.  The spectral region is so full of atomic and
molecular lines that the method of spectrum synthesis must be used.  We have
analyzed our Keck/HIRES spectrum of the sky in the same way.

\begin{figure}[h]
\epsscale{0.6}
\plotone{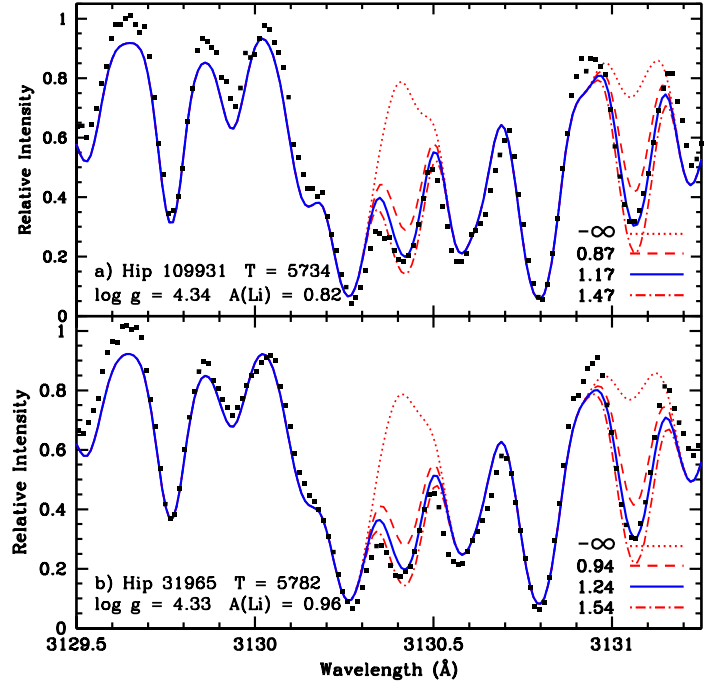}
\caption{Spectrum syntheses in the Be II region for two of our stars.  This
shows 1.75 \AA{} of the 3 \AA{} region that we synthesized.  The observed
spectra are represented by the black squares.  The best fit synthesis is the
solid blue line.  A factor of two more Be is the red dashed-dotted line; a
factor two less Be is the red short dashed line; a fit with no Be is the
red dotted line.}
\end{figure}

Examples of the synthesis fits for two of our stars are shown in Figure 3
along with a synthetic spectrum with no Be at all and one with two times more
Be and with two times less.  Table 3 gives the Be abundances as A(Be), for all
53 stars along with their temperatures and Li abundances as A(Li).  Our value
for A(Be) for the Sun is 1.23.

During the course of the synthesis process, we found two pairs of stars with
virtually identical stellar parameters, but with very different Be contents.
Figure 4 shows the Be II region of the normal Be star, HIP 1813 with A(Be) =
1.01, with that of HIP 32673 whose spectrum is virtually identical in every
feature except the two Be II lines.  The stars differ in temperature by 20 K,
in log g by 0.14, in [Fe/H] by 0.06.  The mass difference is 0.01 and the ages
are similar.  Both are deficient in Li with upper limits on A(Li) of $<$1.08.
Takeda et al.~(2011) also found no evidence for Be in HIP 32673, listing A(Be)
as $<$$-$0.78 and A(Li) as $<$1.04.

\begin{figure}[h]
\epsscale{0.6}
\plotone{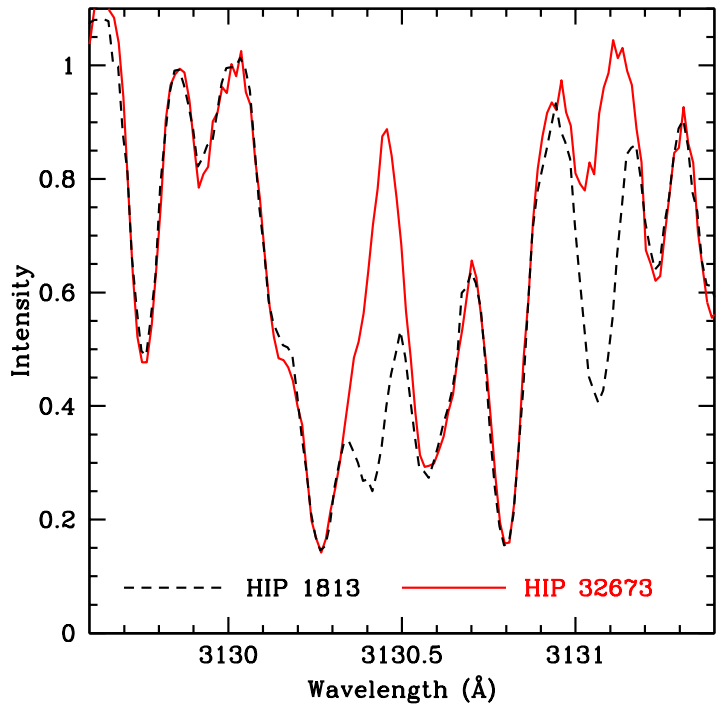}
\caption{Spectra in the Be II region of two similar stars with very different
Be content.  HIP 32673, shown as the solid red line, appears to have no Be
present in either of the two Be II lines.  HIP 1813, shown by the dashed black
line, has A(Be) = 1.01.  The temperature for the Be-less star is 5736 K and
that of HIP 1813 is 5756 K.  Their values of [Fe/H] are solar at 0.03 and
$-$0.03 for HIP 32673 and HIP 1813, respectively and their ages are 6.61 and
7.95 respectively.}
\end{figure}

Figure 5 shows another pair of otherwise identical stars yet one has virtually
no Be: HIP 58576 with normal Be at A(Be) = 1.21 and HIP 55846 with no Be.
Those two stars are within 35 K in temperature, within 0.06 in log g, within
0.01 in [Fe/H].  The Li abundance in the Be normal star (HIP 58576) is given
as A(Li) = 0.32, with no upper limit sign; for HIP 55846, with undetectable
Be, A(Li) is listed as 0.96, again, not listed as an upper limit (Ramirez et
al.~2012)).

The properties of these two pairs of stars along with their Li and Be
abundances are shown in Table 4.

\begin{figure}[h]
\epsscale{0.6}
\plotone{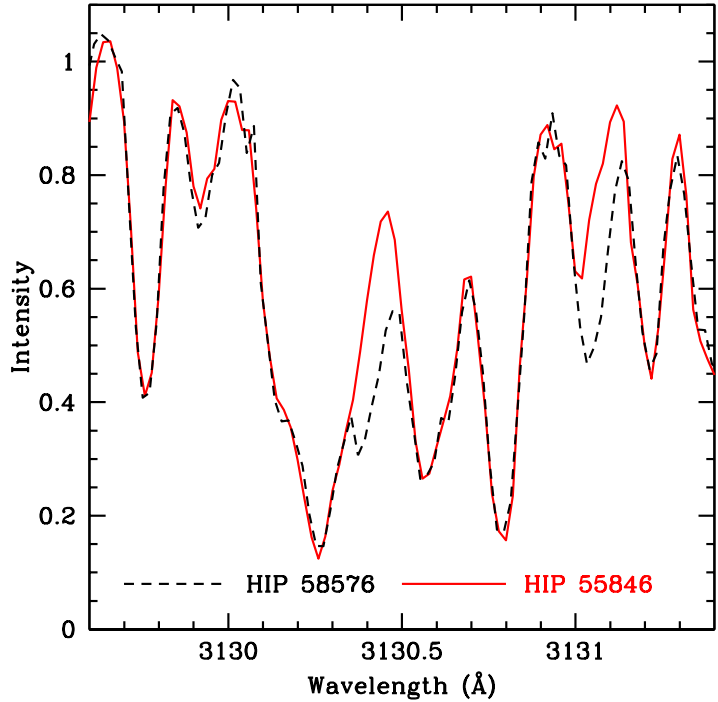}
\caption{Spectra in the Be II region of two similar stars with very different
Be content.  HIP 55846, solid red line, appears to have no Be present in
either of the two Be II lines.  The black dashed line shows the Be region of
HIP 58576 with A(Be) = 1.21.  As in the previous figure these two stars are
 similar in other respects: HIP 55846, with no Be has T = 5463 K, [Fe/H] =
0.30, age = 3.45 Gyr, while HIP 58576, dashed black line, has T = 5508 K,
[Fe/H = 0.31, age = 6.71 Gyr.}
\end{figure}

These two stars, HIP 32673 and HIP 55847, with no apparent Be II lines do show
evidence of the presence of lines that blend with that Be II line.  That
spectral feature is due to a blend of Mn II at 3131.015, Mn I at 3131.037 and
Fe I at 3131.043 \AA.  Those lines are included in our synthesis list.

\subsection{NLTE Discussion}

Calculations have been made of non-local thermodynamical effects (nLTE) on the
Be abundance.  A complete analysis was done by Chmielewski, Mueller, \& Brault
(1975) of those effects in the solar spectrum.  They found that Be is more
ionized than expected in LTE and that the metastable level, 2p$^3$P$^0$, of Be
I is underpopulated.  They calculated that the effects in the Sun were
$\leq$0.10 dex.  Those effects, caused by a hot radiation field in the uv, had
been found to cancel out for Be by Kiselman \& Carlsson (1995) and to be less
than 0.10 dex.  These nLTE investigations were extended to more metal-poor
stars by Garcia Lopez, Severino, \& Gomez (1995).  They also conclude that the
effects are $<$0.10.  Additional discussion about nLTE effects can be found in
the review article by Asplund (2005).

Very recently, Korotin \& Kucinskas (2022) used new atomic data and calculated
and tabulated the corrections for nLTE effects on Be for four values of
[Fe/H]: $-$2.0, $-$1.0, 0.0 and +0.5 for three values of [Be/Fe]: $-$0.5, 0.0
+0.5.  Their calculations cover six temperatures from 4500 - 6500 K and
seven values of log g from 0.0 to 5.0.  For stars with our solar parameters the
effects are small, e.g. at T = 6000 K, log g = 4.5, [Fe/H] = 0.0, the
correction is -0.06 dex.  We have not applied this correction to our results.

\subsection{Error Discussion}

Ramirez et al.~(2012) list the errors in the stellar parameters for all 1381
1381 stars in their Li abundance study.  In turn we have examined those errors
in $T_{\rm eff}$, log g, and [Fe/H] in our subset of those stars.  Virtually
all show $\pm$50 K in temperature, and $\pm$0.04 in [Fe/H] with a range of
$\pm$0.03 to $\pm$0.08 in log g.  In all of our Be analyses we have evaluated
Be errors in grid model atmospheres for 2 values of Be line blends for four
temperatures (5750, 6000, 6250, 6500 K), three log g values (4.5, 4.0, 3.5),
three metallicities ([Fe/H] = $-$0.1, 0.00, +0.1) and two microturbulent
velocities (1.25 and 1.5 km sec$^{-1}$).  From that we can assess the errors
in the Be abundances which are due to the choice of those stellar parameters.
For example, for an error in log g of 0.5, the error in A(Be) was found to be
$\pm$0.25.  For our stars the typical error in log g from Ramirez et
al.~(2012) is 0.05 so that would result in an error in A(Be) of $\pm$.025.  An
uncertainty of $\pm$50 K in temperature results in an uncertainty of $<$0.01
in A(Be).  The uncertainty of 0.05 in [Fe/H] gives a Be abundance uncertainty
of $\pm$0.02.

The uncertainty in A(Be) does not only result from the errors associated with
the stellar parameters.  The ultraviolet spectral region where the Be II
resonance lines are located is very crowded with atomic and molecular lines.
The identification of these lines as well as information on the excitation
potentials and especially the transition probabilities are input into the
abundance determination.  The list of blending lines, their wavelengths, and
their atomic characteristics can be another potential source of error even
though those are well-known for the two Be II lines.

The model atmosphere and the line list provide calculated spectra; the process
of matching that with the observed has a subjective element.  This introduces
another uncertainty into the final value of the Be abundance which is
difficult to quantify.  Overall, we suggest the the typical error in A(Be) is
$\pm$0.12.

We have results for nine stars to compare with those of Takeda et al.~(2001)
so we can compare our parameters and abundances of Li and Be with their
values.  There is a systematic difference in that all our Be abundances are
lower than theirs with a range of $-$0.01 to $-$0.35 and a mean difference of
$-$0.17 $\pm$0.12.  Our temperatures are in good agreement; on average ours
are $\sim$12$^{\circ}$ K hotter.  Our values for log g are within $\pm$0.03 on
average.  We suspect the Be differences are therefore contained in the line
list parameters we use in our respective syntheses.  In the region between
3129.0 and 3132.9 \AA{} our list does contain 171 lines compared to theirs of
124 or 47 more lines than their list.  Many weak lines would tend to depress
the continuum in the calculated spectrum and lower the Be abundance.

\section{RESULTS}

Our final results for A(Be) are given in Table 3 along with the values for
A(Li) from Ramirez et al.~(2012) for 53 stars.  The upper limits on the values
we found for Be in two stars and those reported by Ramirez et al.~(2012) for
Li in 12 stars are indicated by $<$ signs.  (Although we observed and analyzed
HIP 394, we have not included it in any of the figures because it is a
subgiant and out of our range in mass and metallicity.)

\begin{figure}[htb!]
\epsscale{1.0}
\plottwo{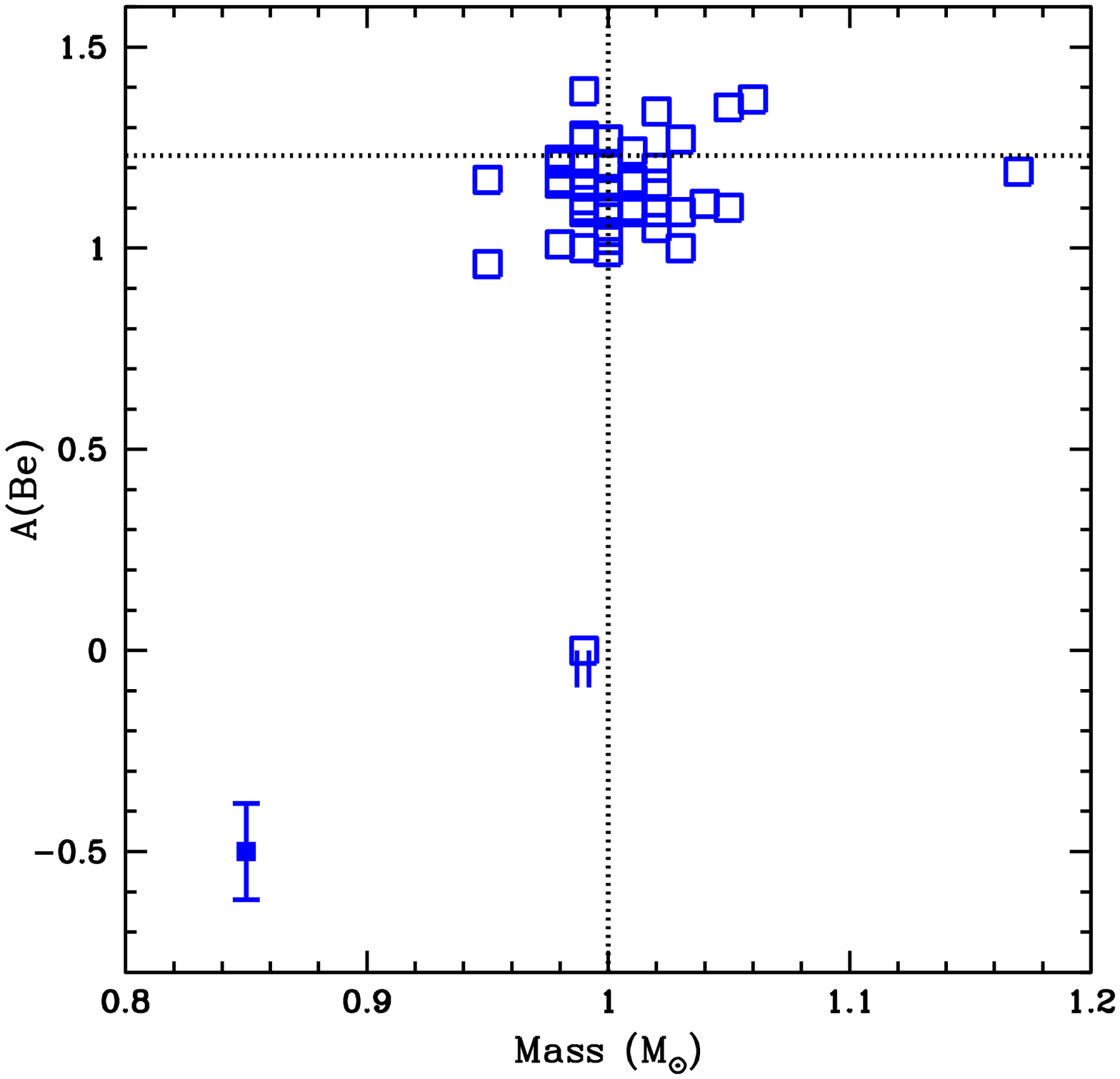}{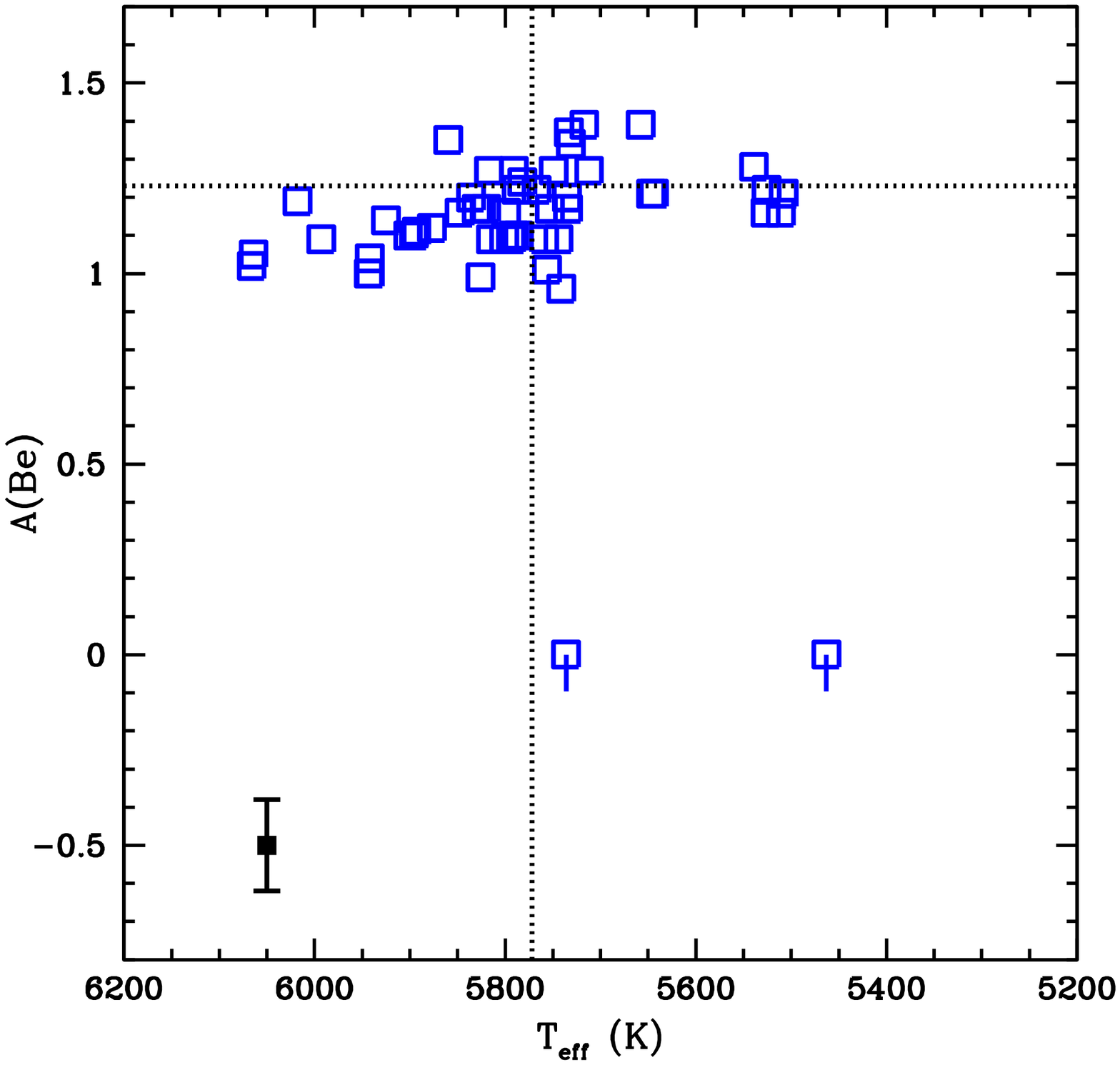}
\caption{Left: Beryllium abundances, A(Be), as a function of stellar mass
shown as open blue squares.  Right: Beryllium abundances, A(Be), as a function
of surface temperature.  For each figure the short vertical blue lines
attached to two of the open squares indicate upper limits on the Be abundance.
The horizontal dotted line represents the solar Be abundance at A(Be) = 1.23
from our Keck/HIRES sky spectrum.  The vertical blue dotted lines indicate the
solar mass and temperature, respectively.  A typical error bar is shown in the
lower left.}
\end{figure}
 
In Figure 6 we show our results for the Be abundances with mass (left) and
with temperature (right).  The horizontal line in each graph represents the
solar Be abundance at A(Be) = 1.23 derived from our Keck/HIRES sky spectrum.
The solar mass value is indicated by the vertical line and the solar
temperature of 5772 K is the vertical line in that plot.  The error bar for
A(Be), $\pm$0.12, is shown in the lower left for each graph.  The spread in
A(Be) is larger than a typical one sigma error bar, but within 3 sigma.  About
80\% of our sample is within 2\% of the solar mass with a mean value of A(Be)
of $\sim$1.2.  There is no theoretical expectation of any trend of A(Be) with
temperature.

\begin{figure}[h]
\plottwo{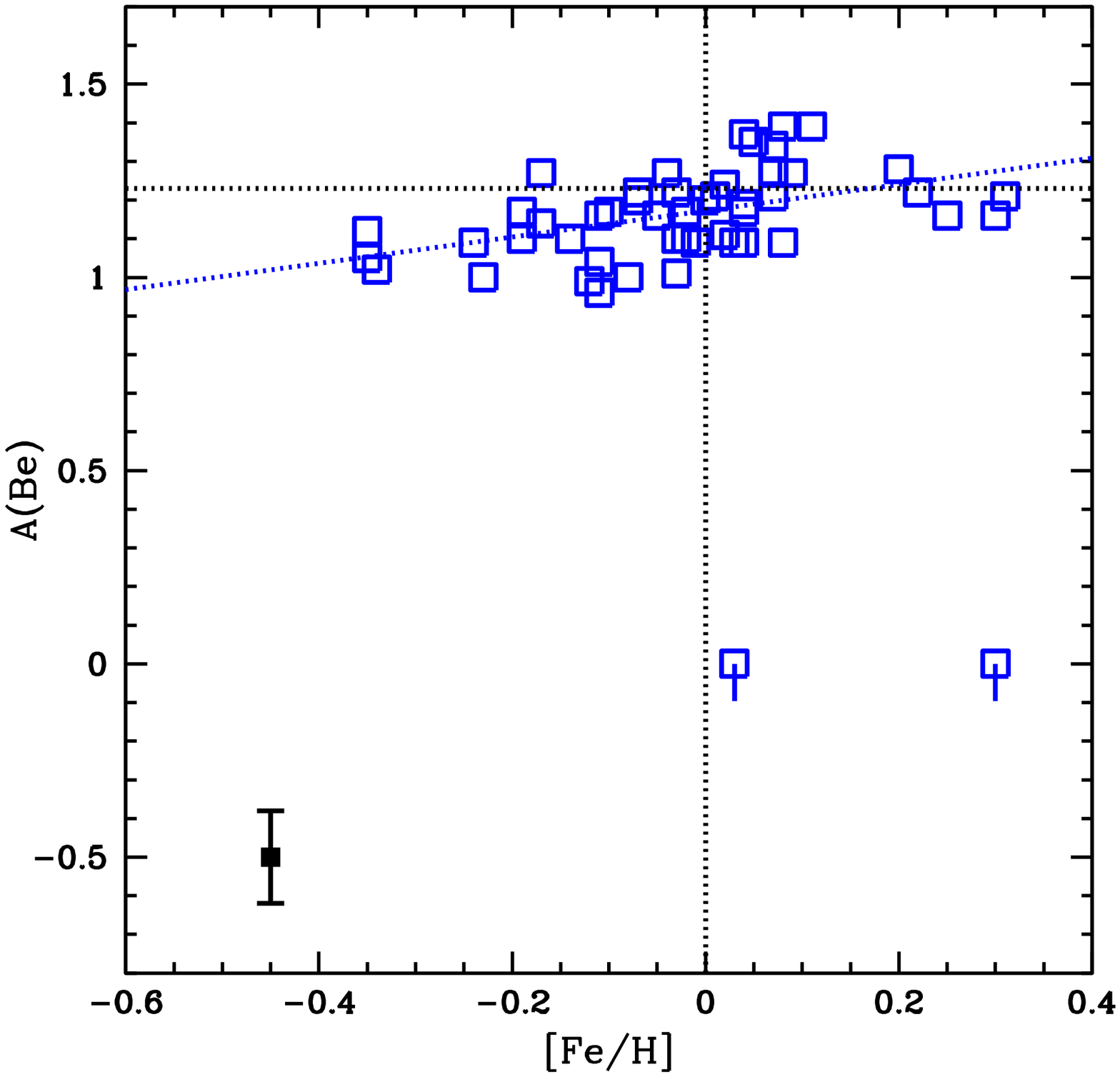}{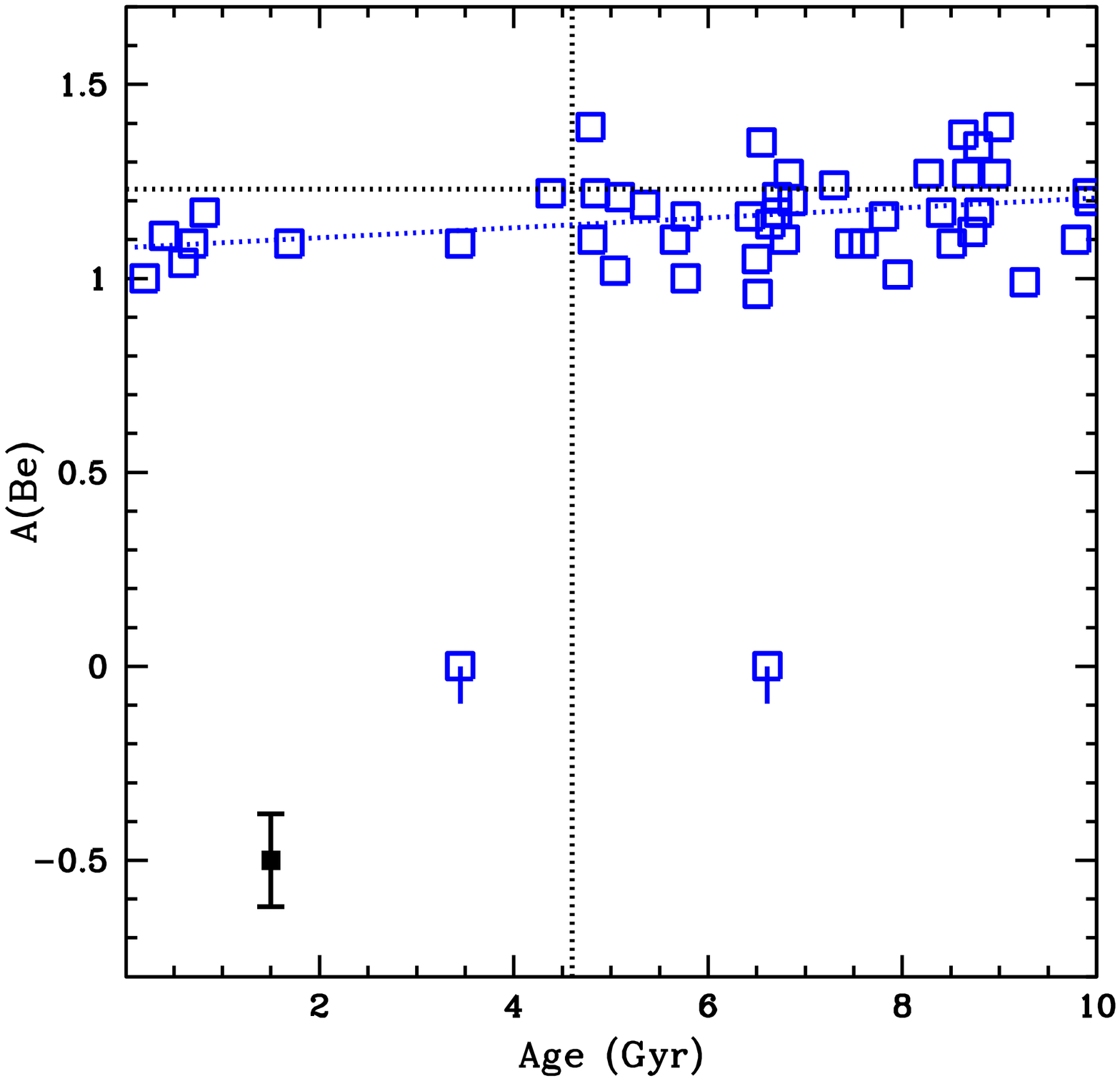}[h]
\caption{Left: Beryllium abundances, A(Be), as a function of [Fe/H] shown as
open blue square.  The blue dotted line through the points is a least-squares
fit with a slope of 0.34 $\pm$0.09.  Right: Beryllium abundances, A(Be), as a
function of age. The blue dotted line is a least-squares fit through the
points and has a slope of 0.013 $\pm$0.005.  For both figures the vertical
blue lines attached to two points indicate upper limits on the Be abundance
and were not included in the fit. The horizontal dotted line represents the
solar Be abundance at A(Be) = 1.23 our Keck/HIRES sky spectrum.  The vertical
dotted line represents the solar mass and temperature, respectively.  A
typical error bar is shown in the lower left.}
\end{figure}

Figure 7 shows our Be results with [Fe/H] (left) and with stellar age (right).
There is a trend of A(Be) with [Fe/H] in this small range in [Fe/H].  This is
similar to that found by Boesgaard et al.~(2004) for 20 solar-like stars with
temperatures spanning 5618 -- 6718 K and metallicities from [Fe/H] of $-$0.52
to +0.11 all having undepleted Be abundances; that measured slope is 0.38
$\pm$0.14.  Our larger sample of 52 stars here gives a slope of 0.34 $\pm$0.09
which is shown in figure 7, left.  It is possible to see a small increase in
A(Be) with stellar age, Figure 7, right.  These would be expected from slow
Galactic production of the rare light elements by cosmic ray spallation and
novae.

One curious result is the identification of two pairs of stars with virtually
identical parameters, but one star in each pair having no detectable Be.
These pairs were shown above in Figures 4 and 5.  We note that Takeda et
al.~(2011) reported four stars with low Be including HIP 32673.  The two stars
with little or no Be, HIP 32673 and 55846, may well have lost their Be, and
Li, through mass transfer or stellar merger with a companion akin to those
discussed by Boesgaard (2007).  In that work the lack of Li and Be in a few
metal-poor halo stars was attributed to destruction of the Li and Be by
thorough and deep mixing in close binaries or merging pairs.  We note that HIP
55846 is a binary star, 83 Leo.  This mass transfer activity may also account
for the other 3 stars reported by Takeda et al.~(2011) to have no Be.

\begin{figure}[htb!]
\epsscale{0.5}
\plotone{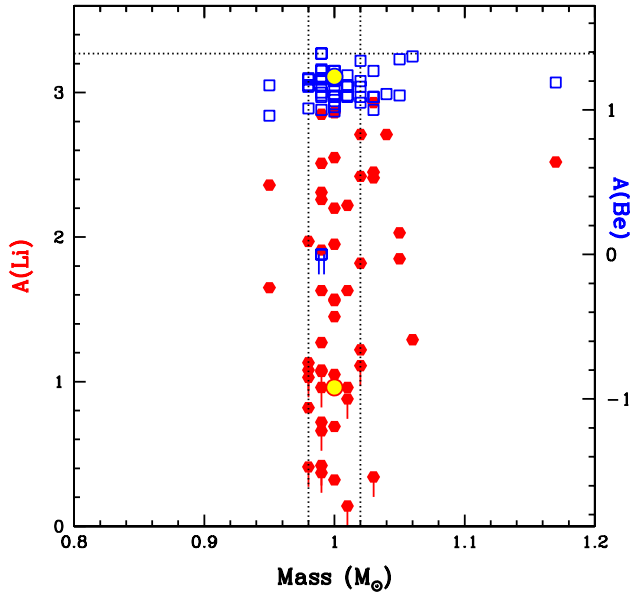}
\caption{Abundances of Li and Be as a function of stellar mass.  The
abundances of the two elements are on the same scale, normalized to their
respective undepleted abundances: A(Li) = 3.27 from meteorites (Lodders (2021)
and A(Be) = 1.38 for the Sun (Asplund, Amarsi \& Grevesse (2021).  The values
for A(Li) are red hexagons and for A(Be) are blue open squares.  Upper limits
on the abundances of both elements are indicated by a vertical line beneath
the plotted points.  The solar A(Li) from (Asplund, Amarsi \& Grevesse 2021)
of 0.96 is shown as a filled yellow ball surrounded by red circle.  Our solar
A(Be) value of 1.23 is a yellow ball surrounded by a blue circle.  Whereas the
Li points show a large spread over this small range of stellar mass, the Be
points are clustered together with the exception of the two stars with no
apparent Be.  The two vertical dotted lines delineate the 0.98 and 1.02 solar
masses.}
\end{figure}

In Figures 6 and 7 we have shown our results for Be with our value
for solar Be from our Keck spectrum with a horizontal dotted line at A(Be) =
1.23.  Now we can put our values for Li and Be on the same abundance scale
and normalize them to a common maximum: A(Li) = 3.27 (Lodders 2021) and A(Be)
= 1.38 (Asplund, Amarsi \& Grevesse 2021).  These results are shown in Figures
8 -- 11 for mass, temperature, metallicity, and age.  In each of these figures
our value for solar Be at 1.23 is shown by the large yellow disk inside the
blue circle and the solar value for Li of 0.96 is the large yellow disk inside
the red circle.  In these one solar-mass stars the Li abundances range over
three orders of magnitude while the Be abundances cover a span of little more
than a factor of 2.5 (with the exception of the two stars with only upper
limits on the Be abundance).

\begin{figure}[htb!]
\epsscale{0.5}
\plotone{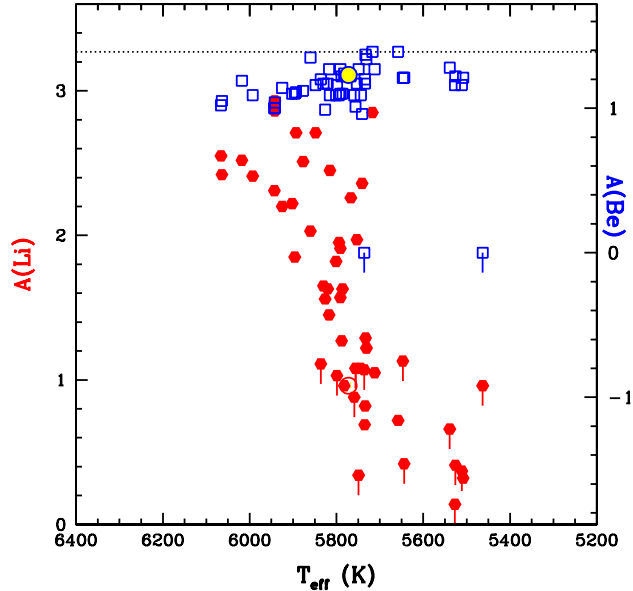}
\caption{Abundances of Li and Be as a function of surface temperature, on the
same scale and normalized as in the previous figure.  The symbols and upper
limits are as in the previous figure as are the solar symbols.  The Li results
show a large spread with a trend toward lower Li with lower temperature.  The
Be results show a minimal spread with two stars having only upper limits.}
\end{figure}

Although we have used the solar Be abundance, A(Be) = 1.38 from
Asplund, Armasi \& Grevesse (2021) to scale our results, this value for the
Sun is not certain.  Those authors note in their discussion of Be that this
number could be too high due to blending features that may not have been
included.  Our Keck daytime sky spectrum yields solar A(Be) = 1.23.  As
mentioned at the end of section 3.2, many weak lines would depress the
continuum and thus lower the Be abundance.

We note that Takeda, Tajitsu, Honda et al.~(2011) found 1.22 for solar
Be.  In their careful and detailed new study of solar Be Carlberg, Cunha,
Smith et al.~(2018) found 1.30 from a spectrum of the asteroid Vesta.  Our
subset of the ten stars that are most similar to the Sun in all five
parameters (see Table 5) have a mean value for A(Be) = 1.18 with a range of
1.09 -- 1.27.  Figure 8 shows the distribution of Li and Be abundances with
stellar mass.  The spread in Li is seen clearly and is particularly true
within the narrower mass span of 0.98 -- 1.02 M$_{\odot}$.  This figure also
highlights the small range of masses in our sample.

The spread in Li abundance with the stellar surface temperature is shown in
Figure 9.  This figure reveals a trend of A(Li) with $T_{\rm eff}$ with more
Li depletion at cooler temperatures with a steep decline from 6000 to 5700 K.
Inasmuch as Be must be mixed deeper to higher temperatures to be destroyed,
there is no such trend seen in the Be abundances.

Figure 10 shows that relationship between stellar metallicity and Be abundance
seen in Figure 7 (left) with a slope of 0.34 $\pm$0.09.  There is a broad
decline in Li with [Fe/H] due to Li depletion.  Although the range in [Fe/H]
is relatively small, a factor of 6, some of the spread may come from the
reduced opacity in the lower metal stars causing reduced mixing and greater Li
retention.  There may be a discernible relation between Be and metallicity as
discussed for Figure 7, left.

\begin{figure}[htb!]
\epsscale{0.5}
\plotone{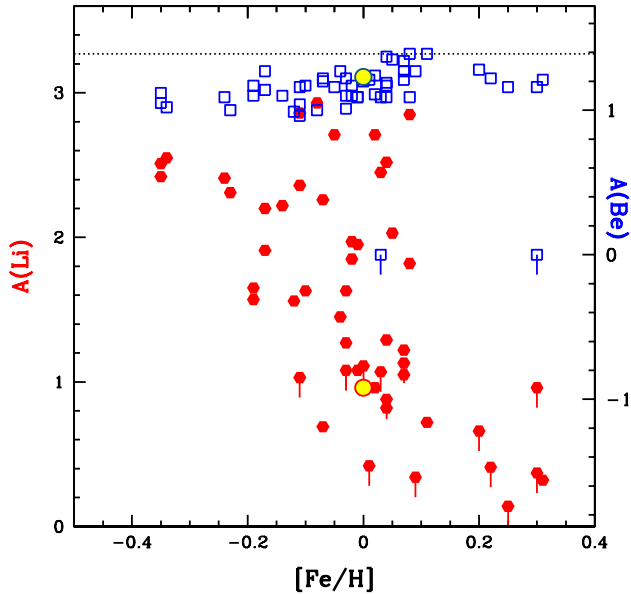}
\caption{Abundances of Li and Be as a function of Fe abundance again
normalized and scaled as in Figure 8.  The symbols and upper limits are as in
the previous figures as are the solar symbols.  There is a general broad
decline in Li with declining metallicity.  For Be there is a mild increase due
to a general galactic production of light elements.}
\end{figure}

Our subset of one solar-mass stars has had age determinations made by Ramirez
et al.~(2012) with isochrones.  (They give error bars which indicate a
sizable range with some of the older stars perhaps 2-3 Gyr.)  Figure
11 shows the Li and Be abundances with stellar age.  There is a spread in Li
at most ages but small Li depletions at the youngest ages.  A trend of
increasing Be with age can be discerned as shown in Figure 7, right.

\begin{figure}[htb!]
\epsscale{0.5}
\plotone{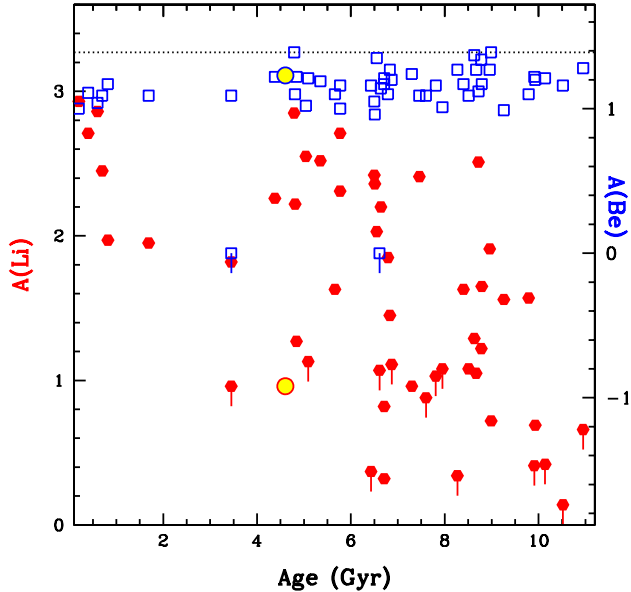}
\caption{Abundances of Li and Be as a function of age, again normalized and
scaled.  The symbols and upper limits are as in the previous figures as are
the solar symbols.  The values of A(Be) show a mild increase with age.
For Li there is a wide spread with age and little noticeable trend with age.}
\end{figure}

We have confined our already restricted sample to the stars that are the most
similar to the Sun in the five properties of mass, temperature, metallicity,
surface gravity and age.  Those stars and the Sun are given in Table 5 along
with their defining characteristics.  The high and low values for mass,
$T_{\rm eff}$, log g, [Fe/H], age and A(Be) are given with the mean values and
probable errors of those means for the ten stars.  The range in A(Be) is 1.09
to 1.27 with a mean of 1.18; however, the spread in A(Li) is much larger: 0.82
to 2.71.

The value of the solar Li is in the lower third of our total sample of solar
mass stars.  In our sample of 10 closest clones of the Sun, however, there is
only one star with less Li than the Sun, HIP 109931, at A(Li) = 0.82.  The Sun
seems exceptionally depleted in Li.  However, it is among the upper third in
its Be content.

Rotational models of light element depletion (e.g.~Deliyannis \& Pinsonneault
1997) show a slope between A(Be) and A(Li) of about 0.4 in F dwarfs in close
agreement with observations (Deliyannis et al.~(1998).  The model slope
decreases with cooler stars as the deeper convection zones play a greater role
in depleting Li but not Be.  We have a subsample of 29 stars between $\pm$100
K of the solar value of 5772 K.  The range of Li/Be is 0.31 to 35.5.  Figure
12, left, shows the plot of A(Li) with A(Be) for those stars showing a slope
of $-$0.008, virtually no slope.  A cleaner subsample is the stars given in
Table 5 that are the 10 most similar to the Sun in all five parameters.  There
are no stars with upper limits on A(Li) in that sample.  For those stars the
Li/Be range is 0.45 to 35.5.  The plot for those stars is shown in Figure 12,
right.  That slope is also negligible at $-$0.022.  The Deliyannis \&
Pinsonneault (1997) models suggest a slope of 0.2 at one solar mass, so either
the models deplete too much Be or other mechanisms contribute to the mixing.
Models show that gravity waves affect Li more than Be (e.g. Garcia Lopez \&
Spruit 1991).

\begin{figure}[htb!]
\epsscale{1.1}
\plottwo{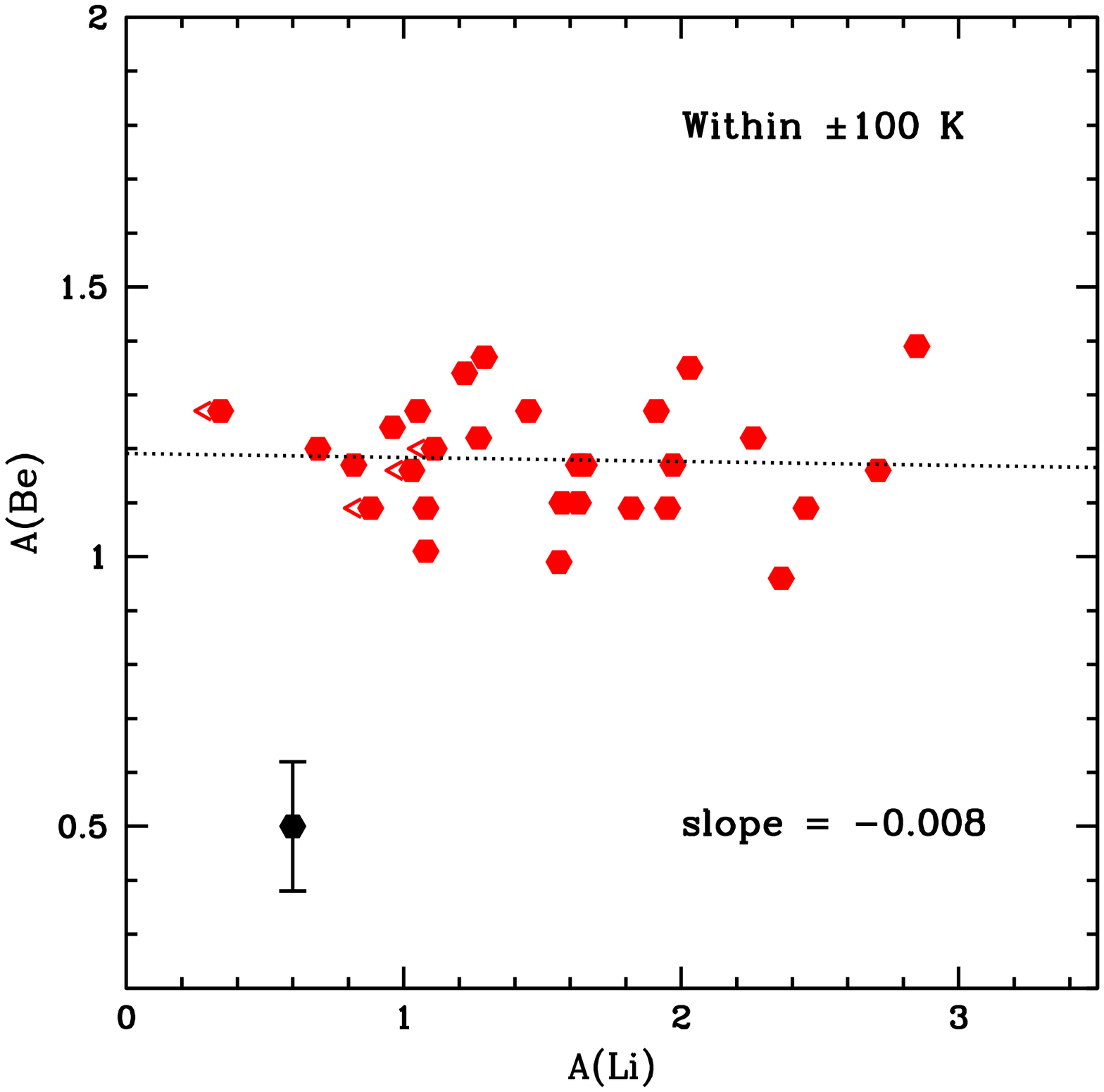}{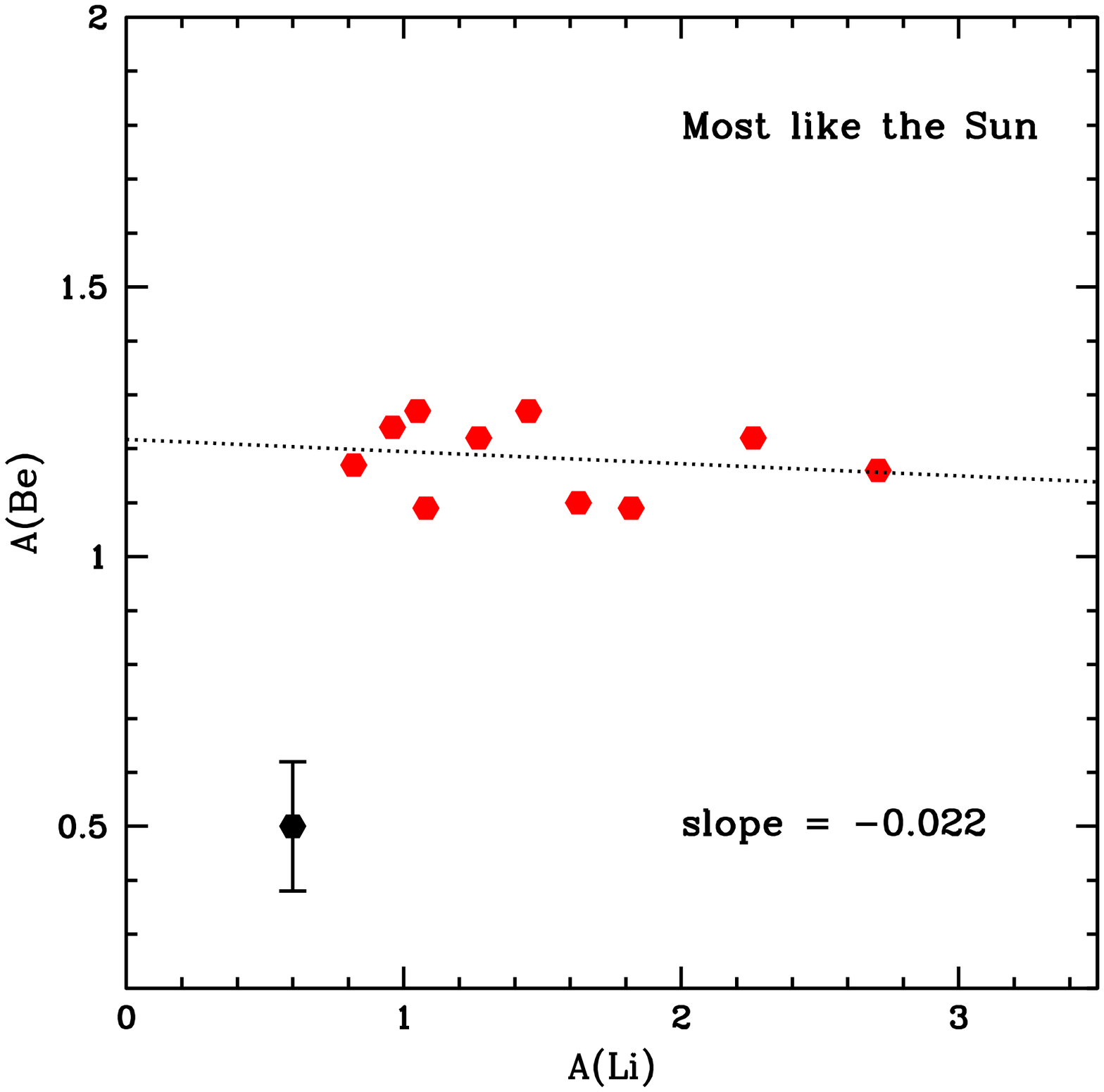}
\caption{Left: Lithium abundance vs.~Beryllium abundance for a sample of stars
within $\pm$100 K of the solar temperature.  Upper limits on A(Li) are shown
with $<$.  There is no discernible slope.  Right: Li and Be abundances for the
10 stars most similar to the Sun from Table 5.  Again the slope is
negligible. }
\end{figure}

\section{SUMMARY AND CONCLUSIONS}

We have made observations of the resonance lines of Be II in 53 main-sequence
stars within approximately $\pm$0.02 M$_{\odot}$ at high spectral resolution
(45,000) and high signal-to-noise ($\sim$50-180) with the upgraded HIRES on
the Keck I telescope.  These observations took place over the course of 4
observing runs between 2014 and 2019.  Our sample comes from stars with known
Li abundances and stellar parameters determined in a uniform way for 1581
stars by Ramirez et al.~(2012).  With those parameters we have found Be
abundances, A(Be), through spectrum synthesis with an advanced version of
MOOG.

We have found two pairs of virtually identical stars in their parameters and
spectra, except one of each pair has normal Be and the other has no evidence
of either Be II line.  We suggest that the stars with no Be, HIP 32673 and
55846, have undergone a mass-exchange event or a merger with a companion that
has thoroughly redistributed matter such that the Li and Be material has been
destroyed at high temperatures by thermonuclear reactions.

When we exclude the two stars with no visible Be features, we find that our
solar-mass stars have very similar Be abundances.  With the exception of the
subgiant star, HIP 394, the spread in A(Be) is 0.96 to 1.39 = 0.43,
corresponding to a factor of 2.7 for these G dwarfs.  We estimate the error in
the determination of A(Be) to be $\pm$0.12.

For this set of stars the range in A(Li) found by Ramirez et al.~(2012), is
$<$0.14 to 2.94 or $\>$2.80 corresponding to over 630 times.  Considering
only the stars with detectable Li the range is 2.62 or a factor of more than
400.

Standard Li depletions corresponding to our solar-mass stars occur only during
the pre-main sequence evolution (Deliyannis, Demarque \& Kawaler 1990),
Pinsonneault 1997) when the stars are cooler and are fully convective.
Observations of the young Pleiades show no Li depletion among the F and G
dwarfs (Pilachowski, Booth \& Hobbs 1987; Boesgaard, Budge \& Ramsey, 1988).
This is matched by calculations of Li-temperature relation for slowly-rotating
young stars (Somers \& Pinsonneault 2014).  However, Li abundances in main
sequence stars in older open clusters indicate that Li depletion continues
during the main sequence phase.  Consistent with this, our entire sample lies
below the Pleiades Li value of A(Li) $\geq$3.0.  This requires an additional
mixing mechanism(s) and rotational mixing is a prime candidate (Deliyannis \&
Pinsonneault 1997; Somers \& Pinsonneault 2014).

We have shown our Be abundance results as a function of mass and temperature
in Figure 6 and as a function of metallicity and age in Figure 7.  There is a
linear relationship showing that A(Be) increases with [Fe/H] and with age
which would be expected from galactic enrichment of the rare light elements
given no stellar depletion losses; such depletion is not  expected for Be in
solar-mass stars.

The large range in Li compared to the narrow range in Be can be seen
especially clearly in our Figure 8 where those abundances are displayed with
stellar mass.  The abundance results are plotted on the same scale and
adjusted to their respective solar/solar system values.  In stellar mass, most
of our sample of stars are between 0.98 and 1.02 M$_{\odot}$.

In the narrow range of mass, 0.98 -- 1.02 we find a small dispersion in A(Be)
with a range from 0.99 to 1.39.  However, Ramirez et al.~(2012) found the
spread in A(Li)  in these stars covered 0.32 to 2.94 as shown in Figure 8.

Figure 9 shows a steep decline of A(Li) with $T_{\rm eff}$ such that the
cooler stars show greater amounts of Li depletion than the warmer ones.  The
surface amount of Be seems unaffected in these stars.  There is considerable
spread in the Li abundances at any temperature.  Some may be intrinsic spread
and errors, but some can be attributed to age and metallicity differences and
some may be inherent in the depletion mechanisms.  The limits we can place on
Be depletion provide more constraints.  The rotational mixing models of
Deliyannis and Pinsonneault (1997) deplete Be by as much as 0.4 dex.  The
total range for our solar mass stars is 0.4 dex; some may be beyond the errors
and be intrinsic.  If our errors are 0.12 dex, then any intrinsic scatter must
be smaller.  This limit also constrains the efficiency of mixing below the
surface convection zone, and may allow mixing by gravity waves to play a role.
Compared to models with rotational mixing, models with mixing due to gravity
waves affect Li more than Be.

There is a trend of A(Be) with [Fe/H] as seen in Figure 7 (left) and Figure
10.  For our 52 stars with [Fe/H] between $-$0.4 and +0.4 we have found a
slope of 0.34 $\pm$0.09.  This is a good indicator of the gradual production
of the light elements during galactic evolution by cosmic rays spallation and
in novae.  Such a trend can not be delineated well with Li as it is so readily
depleted by stars.  This effect in Be with stellar age is seen also, but ages
are not as well-determined as are [Fe/H] values.  There is a large spread in
Li over the 11 Gyr age range of our stars revealed in Figure 11.

We have looked at the 10 stars most similar to the Sun in all five parameters:
mass, metallicity, age, effective temperature, and surface gravity.  For those
10 stars we have found that the range in A(Li) is 0.82 - 2.71, a factor of 80,
while the range in A(Be) is only 1.09 - 1.27, a factor of 1.5.  The Li/Be
ratio spans 0.45 to 35, almost 2 orders of magnitude, due to the range in Li
abundance.  Figure 12 shows there is no trend between Li and Be abundances in
that subsample nor in the sample of 29 stars within $\pm$100 K of the solar
temperature.  This result suggests that additional mechanisms may contribute
to the mixing of Li below the convection zone.

Our selection of stars from a large and uniform collection FGK stars observed
for Li provides a means to investigate an aspect of stellar interiors and
evolution.  The inclusion of Be abundances gives additional depth to the
investigation. 

\acknowledgements We wish to express our appreciation for their knowledgeable
help during our observing runs to the Keck Observatory support astronomers and
staff.  A.C. acknowledges her graduate research fellowship from the National
Science Foundation (DGE18422402).  C.P.D. is grateful for support through the
National Science Foundation grant AST-1909456.

Facilities: Keck I: HIRES

Software: IRAF (Tody, 1986, 1993); MOOG (Sneden 1973, Sneden, Bean, Ivans, et
al.~2012)  

ORCID IDs

Boesgaard  0000-0002-8468-9532

Deliyannis  0000-0003-1125-2564

Lum  0000-0001-7205-1593

Chontos  0000-0003-1125-2564

\clearpage
\begin{deluxetable}{rrclcr}
\tablenum{1}
\tablewidth{0pc}
\tablecaption{Log of the Keck/HIRES Be Observations in Solar Mass Stars}
\tablehead{
\colhead{HIP} & \colhead{HD} & \colhead{V} & \colhead{Date-UT} &
\colhead{Exp-min} & \colhead{S/N} } 
\startdata 
394    & 225239  &   6.10 &  2014 Dec 26 & \phn2 & 102\\
493    & 101     &   7.45 &  2014 Dec 26 & \phn5 & 93 \\
996    & 804     &   8.18 &  2014 Jan 15 &  \phn5 & 71 \\
1411   & 1327    &   9.06 &  2014 Jan 15 & 15 & 65 \\
1813   & 1832    &   8.3  &  2019 Dec 03 & 10 & 156\\
7244   & 9472    &   7.63 &  2014 Dec 27 & \phn7 & 96 \\
7245   & 9446    &   8.35 &  2014 Jan 15 & \phn7 & 85 \\
23627  & 32724   &   7.20 &  2014 Jan 15 & \phn5 & 82 \\
24813  & 34411   &   4.71 &  2017 Nov 10 & \phn1 & 154\\
25002  & 35041   &   7.68 &  2017 Nov 10 & \phn4 & 79 \\
25052  & 34634   &   7.74 &  2017 Nov 11 & \phn7 & 76 \\
25414  & 35073   &   8.34 &  2014 Jan 15 & \phn7 & 44 \\
30860  & 45350   &   7.88 &  2014 Jan 15 & \phn6 & 91 \\
31965  & 47309   &   7.60 &  2017 Nov 11 & 10 & 72 \\
32673  & 49178   &   8.07 &  2014 Jan 15 & \phn7 & 98 \\
35599  & 56196   &   8.96 &  2014 Jan 15 & 17 & 38 \\
41184  & 70516   &   7.70 &  2017 Nov 11 & \phn6 & 80 \\
42438  & 72905   &   5.65 &  2014 Jan 15 & \phn5 & 32 \\
42723  & 74156   &   7.61 &  2017 Nov 10 & \phn5 & 108\\
44935  & 78534   &   8.71 &  2014 Jan 15 & 20 & 48 \\
49580  & 87680   &   7.98 &  2014 Dec 28 & 10 & 87 \\
50473  & 80307   &   7.03 &  2014 Dec 27 & \phn4 & 106\\
52153  & 92242   &   8.34 &  2014 Dec 28 & 11 & 90 \\
54196  & 96094   &   7.62 &  2014 Dec 28 & 10 & 104\\
54582  & 97037   &   6.81 &  2014 Dec 27 & \phn4 & 96 \\
55846  & 99491   &   6.49 &  2014 Dec 28 & \phn3 & 85 \\
56572  & 100777  &   8.42 &  2014 Jan 15 & \phn8 & 60 \\
57300  & 102081  &   8.27 &  2014 Dec 28 & 10 & 110\\
58576  & 104304  &   5.54 &  2014 Dec 27 & \phn1 & 78 \\
62039  & 110537  &   7.83 &  2014 Dec 27 & \phn5 & 77 \\
62198  & 110835  &   9.06 &  2014 Jan 15 & 10 & 68 \\
63354  & 112756  &   8.13 &  2014 Dec 27 & \phn5 & 84 \\
65049  & 115968  &   8.02 &  2014 Jan 15 & \phn5 & 72 \\
65708  & 117126  &   7.44 &  2014 Dec 27 & \phn5 & 82 \\
72567  & 130948  &   5.88 &  2014 Jan 15 & \phn3 & 189\\
73146  & 132406  &   8.45 &  2014 Jan 15 & 10 & 56 \\
100017 & 193664  &   5.93 &  2017 Nov 11 & \phn2 & 73 \\
100963 & 195034  &   7.09 &  2017 Nov 11 & \phn3 & 86 \\
100970 & 195019  &   6.97 &  2017 Nov 11 & \phn3 & 95 \\
104075 & 200746  &   7.99 &  2019 Dec 03 & \phn6 & 59 \\
106678 & 205656  &   8.61 &  2019 Dec 04 & 10 & 86 \\ 
107350 & 206860  &   5.95 &  2017 Nov 10 & \phn3 & 170\\
108468 & 208704  &   7.17 &  2017 Nov 10 & \phn3 & 52 \\
109090 & 209858  &   7.79 &  2017 Nov 10 & \phn5 & 122\\
109110 & 209799  &   7.58 &  2017 Nov 10 & \phn5 & 84 \\
109378 & 210277  &   6.63 &  2017 Nov 10 & \phn5 & 155\\
109931 & 24$^{\arcdeg}$4563 &   8.95 &  2017 Nov 11 & 10 & 74 \\
110035 & 211476  &   7.04 &  2019 Dec 03 & \phn6 & 54 \\
112504 & 215696  &   7.33 &  2017 Nov 10 & \phn3 & 47 \\
115370 & 220255  &   7.77 &  2017 Nov 11 & \phn6 & 108\\
116906 & 222582  &   7.69 &  2017 Nov 11 & \phn6 & 50 \\
118115 & 224383  &   7.88 &  2017 Nov 11 & \phn8 & 48 \\
118159 & 224448  &   9.01 &  2017 Nov 11 & 12 & 77 \\
\enddata
\end{deluxetable}

\clearpage
\begin{deluxetable}{rcccrrcc}
\tablenum{2}
\tablewidth{0pc}
\tablecaption{Stellar Parameters for the Solar Mass Stars}
\tablehead{
\colhead{HIP} & \colhead{$T_{\rm eff}(K)$} & \colhead{log g} 
& \colhead{[Fe/H]} & \colhead{$\xi$} & \colhead{A(Li)} & \colhead{Mass} 
& \colhead{Age Gyr} }
\startdata 
394    & 5636 & 3.76 & $-$0.49 & 1.92  & 1.98 & 1.10 & 5.00  \\
493    & 5943 & 4.38 & $-$0.23 & 1.36  & 2.31 & 0.99 & 5.77  \\
996    & 5817 & 4.32 & $-$0.04 & 1.34  & 1.45 & 1.00 & 6.83  \\
1411   & 5788 & 4.45 & $-$0.03 & 1.14  & 1.27 & 0.99 & 4.84  \\
1813   & 5756 & 4.30 & $-$0.03 & 1.32  & $<$1.08& 0.98 & 7.95  \\
7244   & 5767 & 4.46 & $-$0.07 & 1.12  & 2.26 & 0.99 & 4.38  \\
7245   & 5801 & 4.42 & $+$0.08 & 1.20  & 1.82 & 1.02 & 3.45  \\
23627  & 5791 & 4.15 & $-$0.19 & 1.54  & 1.57 & 1.00 & 9.79  \\
24813  & 5860 & 4.23 & $+$0.05 & 1.48  & 2.03 & 1.05 & 6.55  \\
25002  & 5741 & 4.44 & $-$0.11 & 1.12  & 2.36 & 0.95 & 6.51  \\
25052  & 5749 & 4.24 & $+$0.09 & 1.39  & $<$0.34 & 1.03 & 8.27  \\
25414  & 5647 & 4.46 & $+$0.07 & 0.98  & $<$1.13 & 0.98 & 5.09  \\
30860  & 5527 & 4.18 & $+$0.25 & 1.29  & $<$0.14 & 1.01 &10.52  \\
31965  & 5782 & 4.33 & $+$0.02 & 1.30  & 0.96 & 1.01 & 7.30  \\
32673  & 5736 & 4.44 & $+$0.03 & 1.12  & $<$1.07 & 0.99 & 6.61  \\
35599  & 5820 & 4.19 & $-$0.10 & 1.51  & 1.63 & 1.01 & 8.40  \\
41184  & 5717 & 4.41 & $+$0.08 & 1.14  & 2.85 & 0.99 & 4.79  \\
42438  & 5876 & 4.49 & $-$0.07 & 1.16  & 2.94 & 1.01 & 0.20  \\
42723  & 6018 & 4.11 & $+$0.04 & 1.77  & 2.52 & 1.17 & 5.35  \\
44935  & 5759 & 4.33 & $+$0.04 & 1.28  & $<$0.88 & 1.00 & 7.60  \\
49580  & 5794 & 4.51 & $-$0.01 & 1.07  & 1.95 & 1.00 & 1.69  \\
50473  & 5925 & 4.34 & $-$0.17 & 1.40  & 2.20 & 1.00 & 6.64  \\
52153  & 5791 & 4.15 & $-$0.17 & 1.59  & 1.91 & 0.99 & 8.96  \\
54196  & 5877 & 4.09 & $-$0.35 & 1.68  & 2.51 & 0.99 & 8.72  \\
54582  & 5826 & 4.22 & $-$0.12 & 1.48  & 1.56 & 1.00 & 9.26  \\
55846  & 5463 & 4.48 & $+$0.30 & 0.85  & 0.96 & 0.99 & 3.45  \\
56572  & 5511 & 4.39 & $+$0.30 & 1.00  & $<$0.37 & 0.99 & 6.43  \\
57300  & 6064 & 4.19 & $-$0.35 & 1.70  & 2.42 & 1.02 & 6.50  \\
58576  & 5508 & 4.42 & $+$0.31 & 0.96  & 0.32 & 1.00 & 6.71  \\
62039  & 5658 & 4.34 & $+$0.11 & 1.18  & 0.72 & 0.99 & 8.99  \\
62198  & 5712 & 4.34 & $+$0.07 & 1.23  & 1.05 & 1.00 & 8.67  \\
63354  & 6066 & 4.40 & $-$0.34 & 1.43  & 2.55 & 1.00 & 5.04  \\
65049  & 5539 & 4.25 & $+$0.20 & 1.21  & $<$0.66 & 0.99 &10.95  \\
65708  & 5735 & 4.17 & $-$0.07 & 1.47  & 0.69 & 1.00 & 9.93  \\
72567  & 5942 & 4.39 & $-$0.11 & 1.35  & 2.86 & 1.00 & 0.60  \\
73146  & 5731 & 4.22 & $+$0.07 & 1.40  & 1.22 & 1.02 & 8.78  \\
100017 & 5902 & 4.42 & $-$0.14 & 1.28  & 2.22 & 1.01 & 4.81  \\
100963 & 5786 & 4.44 & $-$0.03 & 1.16  & 1.63 & 0.99 & 5.66  \\
100970 & 5733 & 4.09 & $+$0.04 & 1.57  & 1.29 & 1.06 & 8.63  \\
104075 & 5893 & 4.36 & $+$0.02 & 1.35  & 2.71 & 1.04 & 0.40  \\
106678 & 5644 & 4.14 & $+$0.01 & 1.43  & 0.42 & 0.99 &10.14  \\ 
107350 & 5942 & 4.45 & $-$0.08 & 1.14  & 2.93 & 1.03 & 0.20  \\
108468 & 5799 & 4.34 & $-$0.11 & 1.30  & $<$1.03 & 0.98 & 7.81  \\
109090 & 5993 & 4.23 & $-$0.24 & 1.60  & 2.41 & 1.03 & 7.46  \\
109110 & 5815 & 4.46 & $+$0.03 & 1.15  & 2.45 & 1.03 & 0.70  \\
109378 & 5526 & 4.38 & $+$0.22 & 1.03  & $<$0.41 & 0.98 & 9.91  \\
109931 & 5734 & 4.34 & $+$0.04 & 1.24  & 0.82 & 0.98 & 6.71  \\
110035 & 5830 & 4.35 & $-$0.19 & 1.31  & 1.65 & 0.95 & 8.79  \\
112504 & 5753 & 4.36 & $-$0.02 & 1.23  & 1.97 & 0.98 & 6.82  \\
115370 & 5896 & 4.29 & $-$0.02 & 1.44  & 1.85 & 1.05 & 6.79  \\
116906 & 5744 & 4.31 & $-$0.01 & 1.29  & 1.08 & 0.99 & 8.51  \\
118115 & 5836 & 4.32 & $+$0.00 & 1.36  & $<$1.11 & 1.02 & 6.87  \\
118159 & 5848 & 4.42 & $-$0.05 & 1.23  & 2.71 & 1.02 & 5.77  \\
\enddata
\end{deluxetable}

\clearpage
\begin{deluxetable}{rcrr}
\tablenum{3}
\tablewidth{0pc}
\tablecaption{Lithium and Beryllium Abundances}
\tablehead{
\colhead{HIP} & \colhead{$T_{\rm eff}(K)$} & \colhead{A(Li)} & 
\colhead{A(Be)} 
}
\startdata 
394    & 5636  & 1.98 & 0.79  \\
493    & 5943  & 2.31 & 1.00  \\
996    & 5817  & 1.45 & 1.27  \\
1411   & 5788  & 1.27 & 1.22  \\
1813   & 5756  & $<$1.08& 1.01  \\
7244   & 5767  & 2.26 & 1.22  \\
7245   & 5801  & 1.82 & 1.09  \\
23627  & 5791  & 1.57 & 1.10  \\
24813  & 5860  & 2.03 & 1.35  \\
25002  & 5741  & 2.36 & 0.96  \\
25052  & 5749  & $<$0.34 & 1.27  \\
25414  & 5647  & $<$1.13 & 1.21  \\
30860  & 5527  & $<$0.14 & 1.16  \\
31965  & 5782  & 0.96 & 1.24  \\
32673  & 5736  & $<$1.07 & $<$0.0  \\
35599  & 5820  & 1.63 & 1.17  \\
41184  & 5717  & 2.85 & 1.39  \\
42438  & 5876  & 2.94 & ...  \\
42723  & 6018  & 2.52 & 1.19  \\
44935  & 5759  & $<$0.88 & 1.09  \\
49580  & 5794  & 1.95 & 1.09  \\
50473  & 5925  & 2.20 & 1.14  \\
52153  & 5791  & 1.91 & 1.27  \\
54196  & 5877  & 2.51 & 1.12  \\
54582  & 5826  & 1.56 & 0.99  \\
55846  & 5463  & $<$0.96 & $<$0.0  \\
56572  & 5511  & $<$0.37 & 1.16  \\
57300  & 6064  & 2.42 & 1.05  \\
58576  & 5508  & 0.32 & 1.21  \\
62039  & 5658  & 0.72 & 1.39  \\
62198  & 5712  & 1.05 & 1.27  \\
63354  & 6066  & 2.55 & 1.02  \\
65049  & 5539  & $<$0.66 & 1.28  \\
65708  & 5735  & 0.69 & 1.20  \\
72567  & 5942  & 2.86 & 1.04  \\
73146  & 5731  & 1.22 & 1.34  \\
100017 & 5902  & 2.22 & 1.10  \\
100963 & 5786  & 1.63 & 1.10  \\
100970 & 5733  & 1.29 & 1.37  \\
104075 & 5893  & 2.71 & 1.11  \\
106678 & 5644  & 0.42 & 1.21  \\ 
107350 & 5942  & 2.93 & 1.00  \\
108468 & 5799  & $<$1.03 & 1.16  \\
109090 & 5993  & 2.41 & 1.09  \\
109110 & 5815  & 2.45 & 1.09  \\
109378 & 5526  & $<$0.41 & 1.22  \\
109931 & 5734  & 0.82 & 1.17  \\
110035 & 5830  & 1.65 & 1.17  \\
112504 & 5753  & 1.97 & 1.17  \\
115370 & 5896  & 1.85 & 1.10  \\
116906 & 5744  & 1.08 & 1.09  \\
118115 & 5836  & $<$1.11 & 1.20  \\
118159 & 5848  & 2.71 & 1.16  \\
\enddata
\end{deluxetable}

\clearpage
\begin{deluxetable}{rcccrrcc}
\tablenum{4}
\tablewidth{0pc}
\tablecaption{Two Stars with No Beryllium with Matching Twins with Beryllium}
\tablehead{
\colhead{HIP} & \colhead{$T_{\rm eff}(K)$} & \colhead{log g} 
& \colhead{[Fe/H]} & \colhead{Mass} & \colhead{Age Gyr} & \colhead{A(Li)} 
& \colhead{A(Be)}
}
\startdata 
1813   & 5756 & 4.30 & $-$0.03 & 0.98 & 7.95 & $<$1.08 & 1.01 \\
32673  & 5736 & 4.44 & $+$0.03 & 0.99 & 6.61 & $<$1.07 & $<$0.0 \\
\hline
58576  & 5508 & 4.42 & $+$0.31 & 1.00 & 6.71 & 0.32 & 1.21 \\
55846  & 5463 & 4.48 & $+$0.30 & 0.99 & 3.45 & 0.96 & $<$0.0 \\
\enddata
\end{deluxetable}

\clearpage
\begin{deluxetable}{lccccccc}
\tablenum{5}
\tablewidth{0pc}
\tablecaption{Close Solar Twins}
\tablehead{
\colhead{HIP} & \colhead{Mass} & \colhead{$T_{\rm eff}(K)$} & 
\colhead{log g} &
\colhead{[Fe/H]} &
\colhead{Age (Gyr)} &
\colhead{A(Li)} & 
\colhead{A(Be)} 
}
\startdata
996	&  1.00 & 5817 & 4.32 & $-$0.04 & 6.83 & 1.45 &	 1.27 \\
1411	&  0.99 & 5788 & 4.35 & $-$0.03 & 4.84 & 1.27 &	 1.22 \\
7244	&  0.99 & 5767 & 4.46 & $-$0.07 & 4.38 & 2.26 &	 1.22 \\
7245	&  1.02 & 5801 & 4.42 & +0.08   & 3.45 & 1.82 &	 1.09 \\
31965	&  1.01 & 5782 & 4.33 & +0.02   & 7.30 & 0.96 &	 1.24 \\
62198	&  1.00 & 5712 & 4.34 & +0.07   & 8.67 & 1.05 &	 1.27 \\
100963	&  0.99 & 5786 & 4.44 & $-$0.03 & 5.66 & 1.63 &	 1.10 \\
109931	&  0.98 & 5734 & 4.34 & +0.04   & 6.71 & 0.82 &	 1.17 \\
116906	&  0.99 & 5744 & 4.31 & $-$0.01 & 8.51 & 1.08 &	 1.09 \\
118159	&  1.02 & 5848 & 4.42 & $-$0.05 & 5.77 & 2.71 &	 1.16 \\
Sun/sky &  1.00 & 5772 & 4.44 & 0.00    & 4.603& 0.96 &	 1.23 \\
\hline	
low	&  0.98 & 5712 & 4.31 &	$-$0.07	& 3.45 & 0.82 &	1.09 \\
high	&  1.02 & 5848 & 4.46 &	0.08	& 8.67 & 2.71 &	1.27 \\	
\hline
means	&  1.00 & 5778 & 4.37 & $-$0.002& 6.21 & \nodata & 1.18 \\
p.e.&$\pm$0.01 &$\pm$40 &$\pm$0.05 &$\pm$0.005 &$\pm$1.72 &\nodata & $\pm$0.07 \\
\enddata
\end{deluxetable}

\clearpage

\end{document}